%% file: ObservingPI.tex
\documentclass[aps,prl,twocolumn, groupedaddress]{revtex4-1}
\usepackage[pdftex]{graphicx}
\usepackage{subfigure}
\usepackage{sidecap}

\usepackage{amsmath}
\usepackage{hyperref}

\graphicspath{{pic/}{plot/}}

\include{mevansTex}

\input{abbrev}

\renewcommand{\fixme}[1]{}

\begin{document}


\title{Observation of Parametric Instability in Advanced LIGO}



\def\vup{\vspace{-2pt}}

\author{Matthew Evans} \email{mevans@ligo.mit.edu}
\author{Slawek Gras}
\author{Peter Fritschel}
\author{John Miller}
\author{Lisa Barsotti}
\affiliation{Massachusetts Institute of Technology, Cambridge, MA 02139, USA \vup}

\author{Denis Martynov}
\author{Aidan Brooks}
\author{Dennis Coyne}
\author{Rich Abbott}
\author{Rana Adhikari}
\author{Koji Arai}
\author{Rolf Bork}
\author{Bill Kells}
\author{Jameson Rollins}
\author{Nicolas Smith-Lefebvre}
\author{Gabriele Vajente}
\author{Hiroaki Yamamoto}
\affiliation{California Institute of Technology, Pasadena, CA 91125, USA \vup}

\author{Ryan Derosa}
\author{Anamaria Effler}
\author{Keiko Kokeyama}
\affiliation{Louisiana State University, Baton Rouge, LA 70803, USA \vup}

\author{Carl Adams}         %
\author{Stuart Aston}
\author{Joseph Betzweiser}
\author{Valera Frolov}
\author{Matthew Heinze}  %
\author{Adam Mullavey}
\author{Arnaud Pele}        %
\author{Janeen Romie}
\author{Michael Thomas}  %
\author{Keith Thorne}
\affiliation{LIGO Livingston Observatory, Livingston, LA 70754, USA \vup}

\author{Sheila Dwyer}
\author{Kiwamu Izumi}
\author{Keita Kawabe}
\author{Daniel Sigg}
\affiliation{LIGO Hanford Observatory, Richland, WA 99352, USA \vup}

\author{Stefan Ballmer} 
\author{Thomas J. Massinger} 
\affiliation{Syracuse University, Syracuse, NY 13244, USA \vup}

\author{Alexa Staley}
\affiliation{Columbia University, New York, NY 10027, USA \vup}

\author{Chris Mueller}
\affiliation{University of Florida, Gainesville, FL 32611, USA \vup}

\author{Hartmut Grote}
\affiliation{Max Planck Institute for Gravitational Physics, 30167 Hannover, Germany \vup}

\author{Robert Ward}
\affiliation{Australian National University, Canberra, ACT 0200, Australia \vup}

\author{Eleanor King}
\affiliation{University of Adelaide, Adelaide, SA 5005, Australia \vup}

\author{David Blair}
\author{Li Ju}
\author{Chunnong Zhao}
\affiliation{University of Western Australia, Crawley WA 6009, Australia \vup}


\date{\today}

\begin{abstract}
Parametric instabilities have long been studied as a potentially limiting effect
 in high-power interferometric gravitational wave detectors.
Until now, however, these instabilities have never been observed in
 a kilometer-scale interferometer.
In this work we describe the first observation of parametric instability
 in an Advanced LIGO detector,
 and the means by which it has been removed as a barrier to progress.
\end{abstract}


\maketitle


\def\Pthresh{\bar{P}}
\def\Parm{P_{\rm arm}}

\section{Introduction}
\sslabel{intro}

Opto-mechanical interactions, the moving of masses by radiation pressure,
 are typically of such a tenuous nature that even carefully designed
 experiments may fail to observe them.
In the extreme environment of a high-power interferometric \gw\ detector,
 however, these effects arise spontaneously.
This is true despite the fact that these instruments feature multi-kilogram optics~\cite{Mavalvala2009},
 unlike the micro-gram or nano-gram optics generally used in experiments which 
 target such effects~\cite{Kippenberg2007,Kippenberg2008,McClelland2011,Poot2012}.

Since the seminal work of Braginsky, Strigin, and Vyatchanin in 2001~\cite{Braginsky2001},
 optical parametric instabilities have been extensively studied
 as a potential limit to high-power operation
 of interferometric \gw\ detectors~
 \cite{Braginsky:2002cc,Schediwy:2004ct,
 Ju:2006bl,Ju:2006ip,Ju:2006kc,Gurkovsky:2007hu,
 Strigin:2007kc,Polyakov:2007ds,Gurkovsky:2007hd,Schediwy:2008gv,
 Vyatchanin:2008gh,Zhao:2008ht,Strigin:2008hj,Strigin:2008bl,
 Miao:2008fa,Zhao:2009cv,Meleshko:2009di,Zhang:2010dy,Evans2010,
 Strigin:2010hq,Strigin:2010ic,Liang:2010dc,Gras2010,
 Strigin:2011dy,Heinert:2011ez,Strigin:2012bt,Vyatchanin:2013eg,Danilishin:2014wt,Chen2014}.
These instabilities are of acute interest because, if left unchecked,
 they will limit detector sensitivity by limiting circulating power.

We report on the first observation of a self-sustaining parametric instability
 in a \gw\ detector and the subsequent quenching of this instability.
This observation, at the LIGO Livingston Observatory (LLO)~\cite{httpLIGO}, 
 the culmination of more than a decade of theoretical calculation,
 numerical modeling, and lab-scale experimentation, and serves as confirmation that models which
 have been built to understand the phenomenon of parametric instability are substantially correct.

A variety of techniques have been suggested for overcoming
 parametric instabilities in \gw\ detectors~
 \cite{Degallaix:2007vb,Ju:2008ej,Gras:2008hr,Gras:2009fy,Fan:2010fw,Miller2011,Zhao2015}.
There are a few notable categories of mitigation techniques:
avoid instability by changing the radius of curvature of one or more optics~\cite{Degallaix:2007vb,Zhao2015},
actively damp mechanical modes as they become excited~\cite{Miller2011}, and
prevent instability by increasing the loss of the test mass mechanical modes~\cite{Gras:2006gv,Gras:2008hr,Gras:2009fy,Gras2010}.
The first of these techniques has been demonstrated and found to be effective at LLO.

\begin{figure}[t!]
\centering
\includegraphics[width=0.5\textwidth]{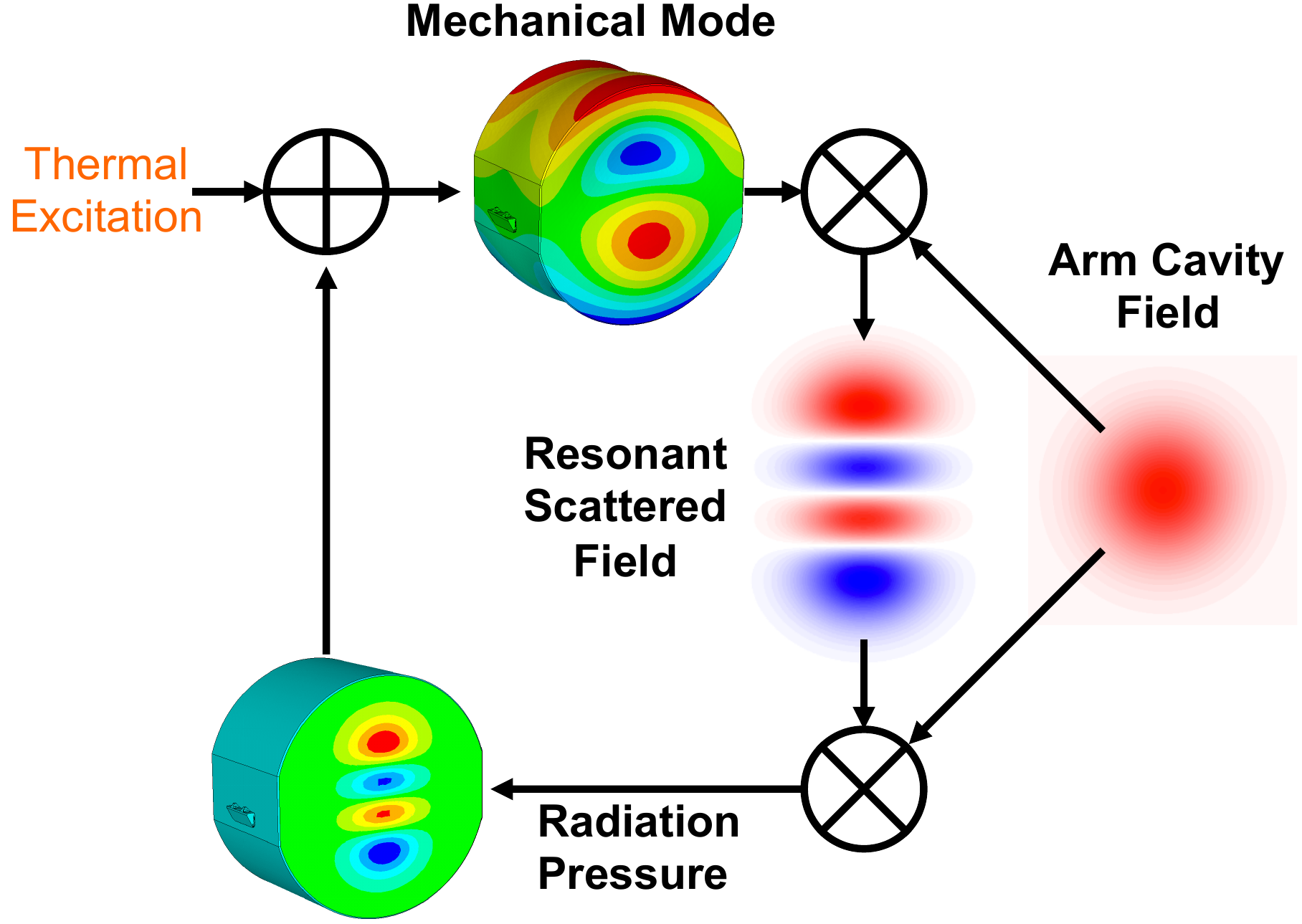}
\caption{Parametric instabilities transfer energy from the arm cavity field
 to a mechanical mode of an interferometer test mass.
Since the rate of energy transfer depends on the amplitude of the excited
 mechanical mode, this can be a runaway positive feedback process.
 \fixme{change modes to match fig 3}
\flabel{LoopPI}}
\end{figure}

\section{Theory of Parametric Instability}
\sslabel{TheoryPI}

Parametric instabilities (PI) operate by transferring energy from the fundamental
 optical mode of the interferometer, which with nearly \SI{1}{MW} of circulating
 power can be as much as \SI{40}{J}, into an interferometer optic's mechanical mode
 (see figure \fref{LoopPI}).
Energy transfer takes place via the radiation pressure driven opto-mechanical interaction
 and the modulation of the fundamental field by the excited mechanical mode~\cite{Evans2010}
 (see figure \fref{LoopPI}).

The parametric gain for a given test mass mechanical mode is given by
\beq{PI_gain}
R_m = \frac{8 \pi Q_m \Parm}{M \omega_m^2 \,c\, \lambda} \sum_{n=0}^\infty \Re[G_n] B_{m,n}^2
\eeq
 (using the notation of reference~\cite{Evans2010}, and correcting an error therein
\footnote{Equation (9) in Evans 2010 is missing a factor of 2,
 resulting from a similar error in equation (24).
This unfortunately matches a factor of 2 error in the equation for $T$
 in section 2 of Braginsky 2001, which should read $T = 4 \pi L / (\lambda_0 Q_{opt})$,
 such that $\delta_1 = (c / 2 L) (2 / T)$.
Note that in Evans 2010 $G_n = 2 / T$
 for Braginsky's Fabry-Perot cavity on resonance (i.e., with $\Delta \omega_n = 0$)}).
The fixed parameters used in equation \eref{PI_gain} are:
 $c$ is the speed of light, $\lambda$ the laser wavelength,
 $M$ the mass of the optic, $\omega_m$ the angular resonant frequency
 of mechanical mode with index $m$, 
 and $B_{m,n}$ the overlap between mechanical mode $m$ and optical mode $n$.

Clearly, the potential for instability grows with increasing power,
 since the parametric gain of all modes is proportional to the power circulating in
 the interferometer arm cavities, $\Parm$.
Active and passive damping techniques for defusing PI operate by reducing 
 the Q of an otherwise unstable mechanical mode, while avoidance techniques
 work by changing the optical gain $G_n$ to prevent instability.

The e-fold growth time for an unstable mode is given by
\beq{PI_tau}
\tau_m =  \frac{2 Q_m}{\omega_m (R_m - 1) }
\eeq
Note that for $R_m = 0$ this gives a negative value equal to the usual decay
 time of a mechanical mode with frequency $f_m = \omega_m / 2 \pi$
 and quality factor $Q_m$.
For values of $R_m > 1$ equation \eref{PI_tau} gives a positive value,
 indicating exponential growth.

The circulating power level $\Parm$ at which a mode becomes unstable
 is referred to as the threshold power $\Pthresh_m$ for that mode,
 and is found by rearranging equation \eref{PI_gain} with $R_m = 1$ and $\Parm = \Pthresh_m$.
At threshold the opto-mechanical interaction is putting 
 energy into the mode at the same rate that it is being dissipated,
 so $\tau_m \rightarrow \infty$.

\section{Observed Parametric Instability}
\sslabel{ObsPI}

Parametric instability was first observed in the Advanced LIGO detector
 recently installed at LLO
 (see reference \cite{Fritschel2014b} for detailed information about the detectors).
The instability grew until it polluted the primary \gw\ output of the detector
 by aliasing into the detection band and saturating detection electronics
 (see figure \fref{piSpectrogram}).
The time scale for growth was long enough to allow for manual intervention 
 (a reduction of the laser power input to the interferometer)
 before control was lost due to this saturation of the readout electronics.

\begin{figure}[t!]
\centering
\includegraphics[width=0.5\textwidth]{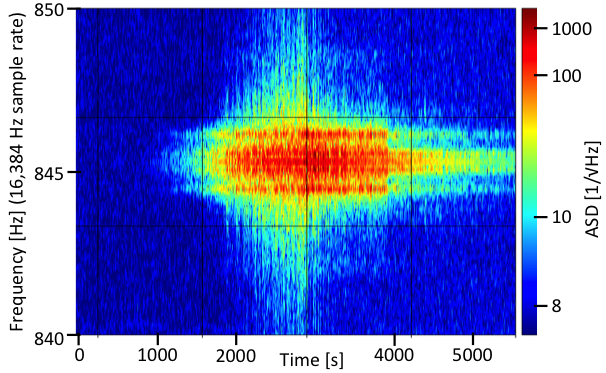}
\vspace{-10pt}
\caption{The excitation due to parametric instability was first observed as
 an exponentially growing feature in the detector's primary output (the \gw\ channel).
The feature is aliased into the detector band by the \SI{16384}{Hz} sample rate
 of the digital system which causes it to appear at \SI{845}{Hz}.
The growing instability eventually causes saturation of the electronic readout chain,
 which appears as broadband contamination of the detector output channel
 (visible between 2500 and \SI{ 3000}{s}).
In the above data, the power into the interferometer was decreased before saturation
 caused the interferometer control systems to fail (see figure \fref{Decay_Power}).
\flabel{piSpectrogram}}
\end{figure}

\begin{figure}[tt!]
\centering
\includegraphics[width=0.45\textwidth]{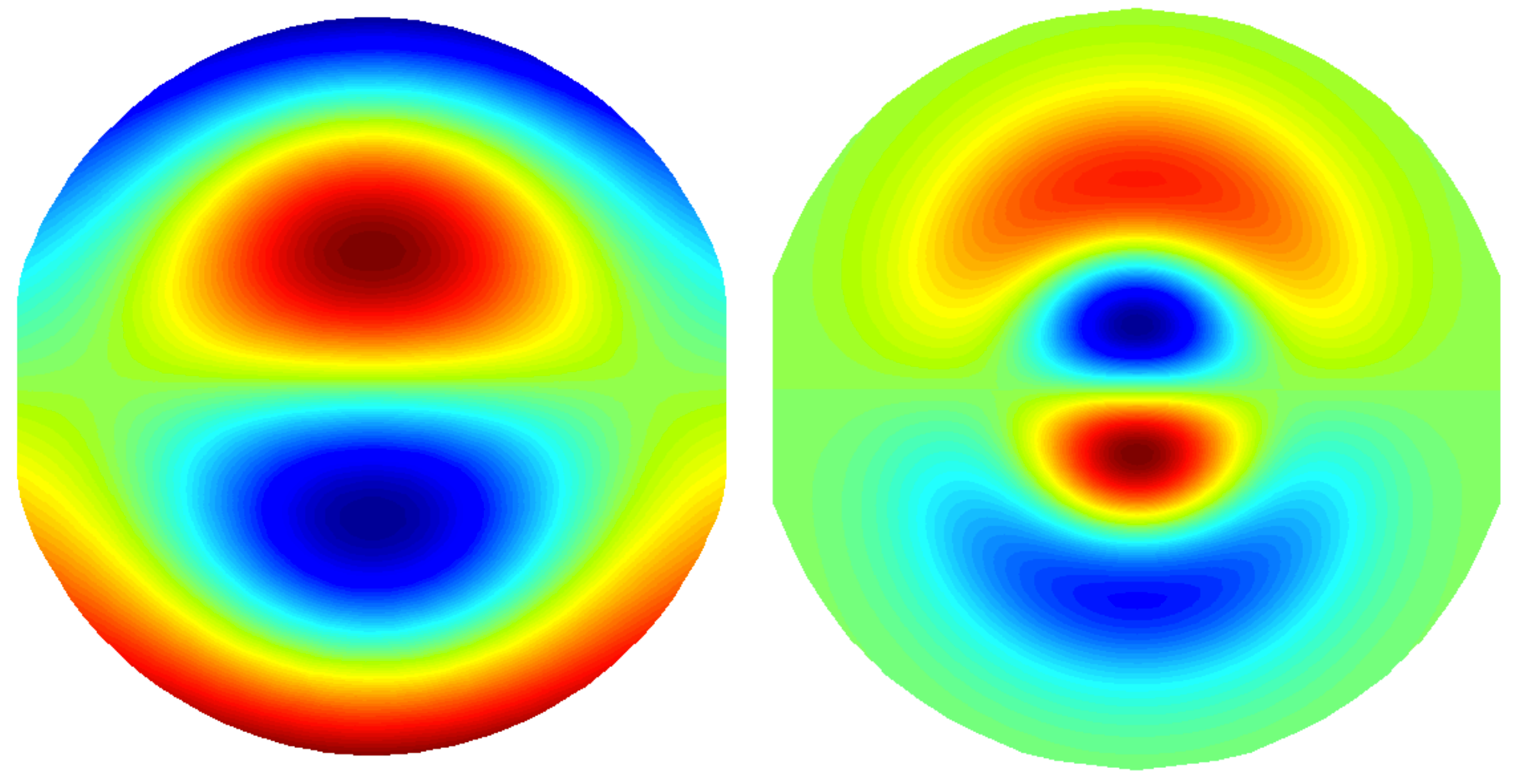}
\vspace{-10pt}
\caption{The test mass mechanical mode and cavity optical mode responsible
 for the observed parametric instability are shown.
The left panel is the test mass front surface displacement due to the mechanical mode,
 while the right panel is the radiation pressure induced displacement
 (red is positive and blue negative in both panels).
Both of these modes occur at \SI{15.5}{kHz} and they have
 an overlap factor of $B_{m,n} = 0.1$.
\flabel{Modes}}
\end{figure}

The test mass (TM) mechanical mode responsible for the instability was
 identified as the \SI{15.54}{kHz} mode shown in figure \fref{Modes}.
The higher-order mode spacing of the Advanced LIGO arm cavities
 is $5.1\pm0.3\,\si{kHz}$, and the optical resonance width is \SI{80}{Hz}, 
 such that a \third\ order transverse optical mode
 can provide energy transfer from the fundamental optical mode to
 this mechanical mode (see figure \fref{LoopPI} and \fref{Modes}).

By measuring the e-folding growth and decay time of the excited acoustic mode
 as a function of circulating power in the interferometer arm cavities,
 we can compute the parametric gain $R$ and mechanical mode quality factor $Q$
 as given in equation \eref{PI_tau}.
Our measurements were performed by operating the interferometer above
 the threshold power to excite the TM mechanical mode
 and then reducing the power below the instability threshold 
 and watching the mode amplitude decrease, as shown in figure \fref{Decay_Power}.

\begin{figure}[t!]
\centering
\includegraphics[width=0.45\textwidth]{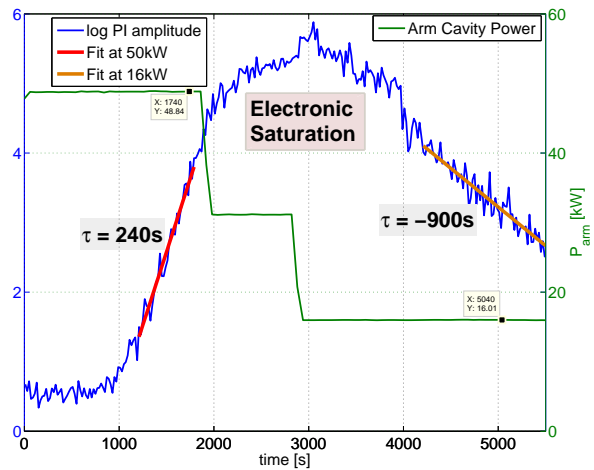}
\vspace{-10pt}
\caption{The amplitude of the excitation shown in figure \fref{piSpectrogram}
 is fitted at times with different values of $\Parm$ to find
 its growth (or decay) time-scale as a function of power.
The growth of the PI is clearly visible above the noise after \SI{1000}{s},
 with an e-folding growth time of \SI{240}{s}, until \SI{2000}{s} at which point
 the readout electronics begin to saturate.
A little more than \SI{4000}{s} into the plotted data, the excited mechanical mode is
 seen decaying with $\tau = -\SI{900}{s}$.
According to equation \eref{PI_tau}, these data imply a threshold
 power of \SI{25}{kW} and $Q = \num{12e6}$
 for this mode.
\flabel{Decay_Power}}
\end{figure}

The observed unstable mode has $R = 2$ with $\Parm = \SI{50}{kW}$,
 and the associated mechanical mode has $Q = \num{12e6}$.
This Q-factor is in the range expected given similar measurements of Advanced LIGO
 test masses~\cite{Miller2011,Gras2015}, and the parametric gain is as predicted by
 equation \eref{PI_gain} for a high, but far from maximal value of $G_n$.
While the 90\% confidence limit for this mode is $R = 11$, as seen in figure  \fref{PI_MaxGain},
 5\% of simulated values are higher than the observed value, and 1\% have $R > 100$.
 

\section{Defusing Parametric Instability}
\sslabel{DefusingPI}

Parametric instability depends on several potentially modifiable features of
 a \gw\ detector, two of which can be taken advantage of in Advanced LIGO
 without modifications to the interferometer core optics.
First, instability requires coincident resonant frequencies of a test mass mechanical mode
 and an arm cavity optical mode
 (i.e., $G_{n}$ in equation \eref{PI_gain} must be large at the mechanical mode
  frequency $\omega_m$ for an optical mode with non-vanishing overlap $B_{m,n}$).

The coincidence of resonant frequencies can be modified by changing the radius of curvature (RoC)
 of one of the test masses in the cavity affected by PI, through its effect on the transverse mode spacing.
Advanced LIGO arm cavities use mirrors with $\sim\!\SI{2}{km}$ RoC, so
 for the \third\ order optical mode shown in figure \fref{Modes}, for instance, a change of \SI{1}{m}
 is sufficient to change its resonant frequency by \SI{80}{Hz}, or one cavity line width.

\begin{figure}[t!]
\centering
\includegraphics[width=0.4\textwidth]{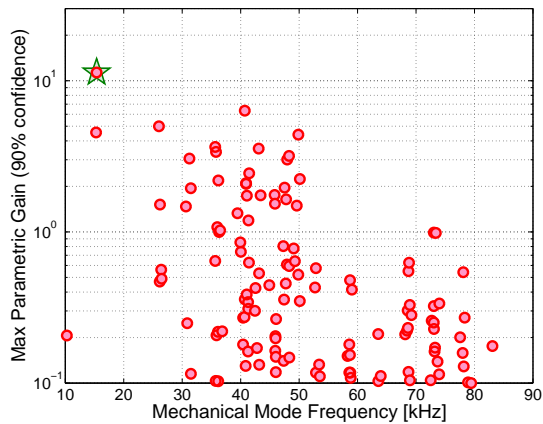}
\caption{The maximum gain expected at full power, $\Parm = \SI{800}{kW}$,
 for all potentially unstable modes (90\% confidence value for a single test mass).
Since the mechanical mode frequencies and optics parameters are known
 with limited precision, the parametric gain of a particular mode cannot be
 computed exactly, and Monte Carlo methods must be used~\cite{Evans2010}.
The ``90\% confidence maximum'' shown here is the value
 for which 90\% of the 342,881 Monte Carlo computations gave a lower gain.
Notably, the observed unstable mode is the mode with the highest predicted parametric gain
 (marked with a green star), and the observed gain is about a factor of 3 higher than the
 value shown here ($R = 2$ at \SI{50}{kW}, or $R = 32$ at \SI{800}{kW}).
This is not so much bad luck as a trials factor; there are 4 test masses, and many potentially
 unstable modes, so it is not unusual that at least one is above the 90\% confidence limit.
In fact, the statistics shown in figure \fref{PI_ProbableNumber} confirm
 that the observation of at least one unstable mode was to be expected.
\flabel{PI_MaxGain}}
\vspace{-10pt}
\end{figure}

A second approach to defusing PI comes from
 the weak radiation pressure coupling of energy from the
 fundamental optical mode to the mechanical modes of the test mass,
 which must be sufficient to overcome the energy loss from the mechanical mode.
This can be seen in the linear dependence of $R_m$ on $Q_m$ in equation \eref{PI_gain}.
The quality factor of mechanical modes in the Advanced LIGO test mass optics
 is roughly $10^7$, and as such is easily spoiled.
Since the highest parametric gain likely to be seen in Advanced LIGO is $\sim\!10$,
 reducing $Q_m$ to less than $10^6$ for potentially unstable modes
 would suffice (see figure \fref{PI_MaxGain}).

In anticipation of PI, all Advanced LIGO test masses were outfitted with electro-static actuators
 capable of damping mechanical modes associated with PI~\cite{Miller2011}.
This method of damping TM mechanical modes was demonstrated at MIT with a prototype
 Advanced LIGO optic, and will be used to damp PI in Advanced LIGO as necessary.

As discussed in the following section,
 the combination of the above techniques will likely prove sufficient if a manageable number
 of unstable modes are encountered.
If, however, the number of unstable modes is very large,
 a broadband reduction of mechanical Q-factors may be required.
Implementing a broadband PI control strategy, as described in~\cite{Gras2015},
 will certainly prove successful, but it may cause some delay in the
 approach of Advanced LIGO to its design operating power.

\begin{figure}[t!]
\centering
\includegraphics[width=0.5\textwidth]{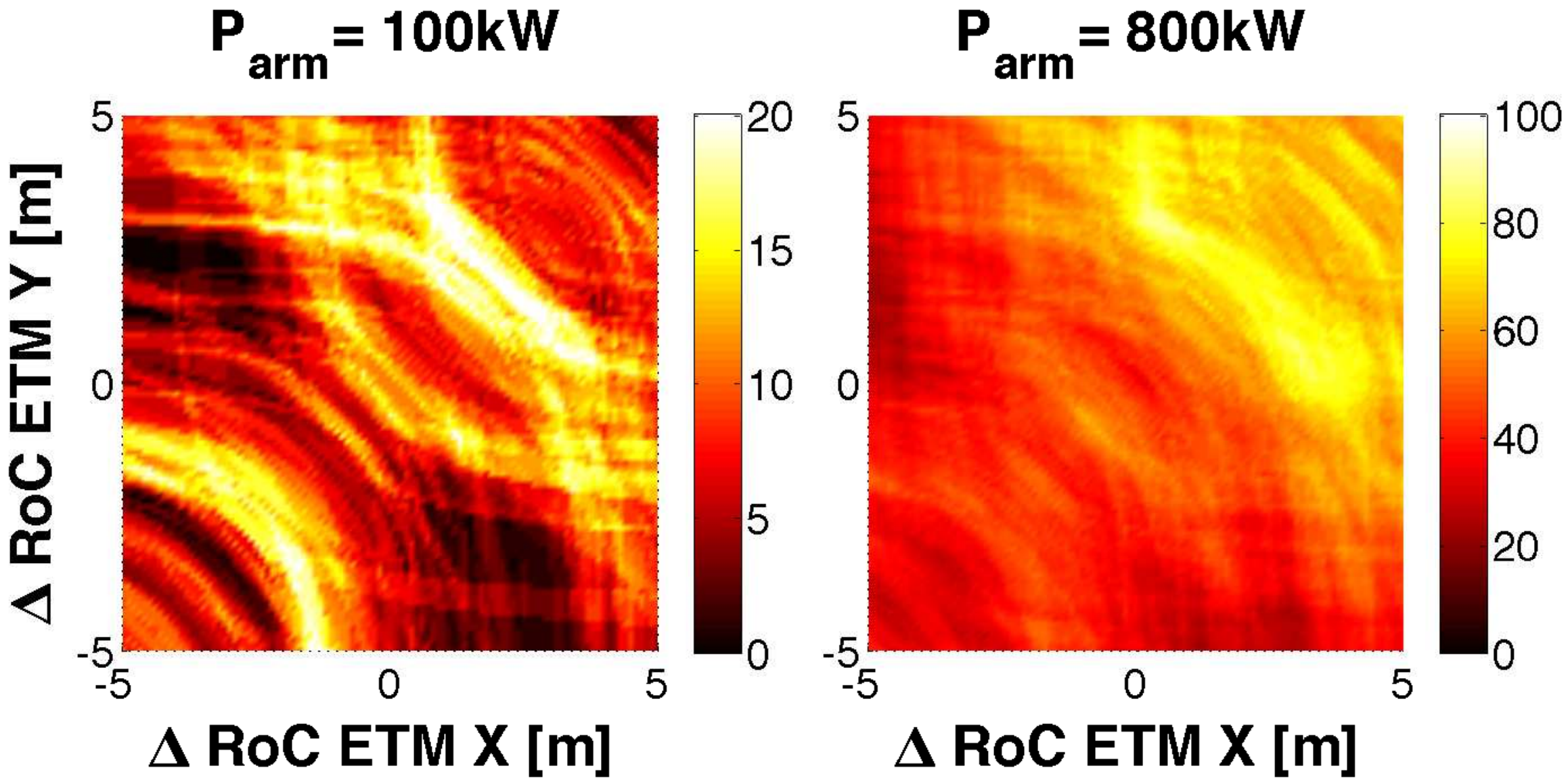}
\caption{The number of unstable modes as a function of change in
 arm cavity mirror radius of curvature (RoC), as indicated by the color scale to the right of each plot.
With \SI{100}{kW} of circulating power, as expected for the first observing run, 
 small changes in mirror RoC can be used to avoid parametric instability.
Higher power levels, however, increase parametric gain; note the different color scales
 used on the left and right panels.
 At Advanced LIGO's design
 operating power of $\Parm = \SI{800}{kW}$ RoC adjustments can at best be used to reduce the
 number of unstable modes. 
\flabel{PI_RH_modes}}
\end{figure}

\section{Implications for the Future of Advanced LIGO}
\sslabel{Implications}

Shortly after the first PI was observed at LLO,
 heating elements, known as ``ring heaters'', were used to change the mirror RoC
  and avoid the instability.
 Ring heaters were included in the Advanced LIGO design to compensate
 for RoC changes due to absorption of the fundamental optical mode,
 and can change the RoC of the optics by several 10s of meters.
After adjusting the ring heaters to produce roughly \SI{2}{m} of RoC change
 of the arm-cavity end mirrors,
 the parametric gain of the observed instability at \SI{15.53}{kHz} was reduced below unity and
  the interferometer was operated with nearly \SI{100}{kW} of circulating power
  for more than 12 hours.

Despite this success, it must be recognized that
 many mechanical modes of the interferometer test masses can be driven to instability;
 figure \fref{PI_MaxGain} shows the maximum expected
 parametric gain of all potentially unstable modes.
From the left panel of figure \fref{PI_RH_modes}, we can conclude that using the
 ring heaters to find a PI-free zone is likely to succeed at present 
 (with $\Parm \sim \SI{100}{kW}$), but the right panel
 tells us that at the design
 value of $\Parm = \SI{800}{kW}$ RoC changes will not be sufficient.

Several major periods of data taking are expected with Advanced LIGO
 at power levels below the design power~\cite{Prospects2013}.
Figure \fref{PI_ProbableNumber} shows the number of modes that
 are likely to be unstable as a function of circulating power in the interferometer.
At the final operating power, for instance, more than 40 modes are likely to be unstable.

In order to operate at the design power level Advanced LIGO will likely need to
 employ multiple methods of defusing PI:
 changing the RoC to reduce the number of unstable modes;
 actively damping the remaining unstable modes with electrostatic actuators;
 and possibly reducing the Q of the test mass  modes with  passive dampers.

\begin{figure}[t!]
\centering
\includegraphics[width=0.45\textwidth]{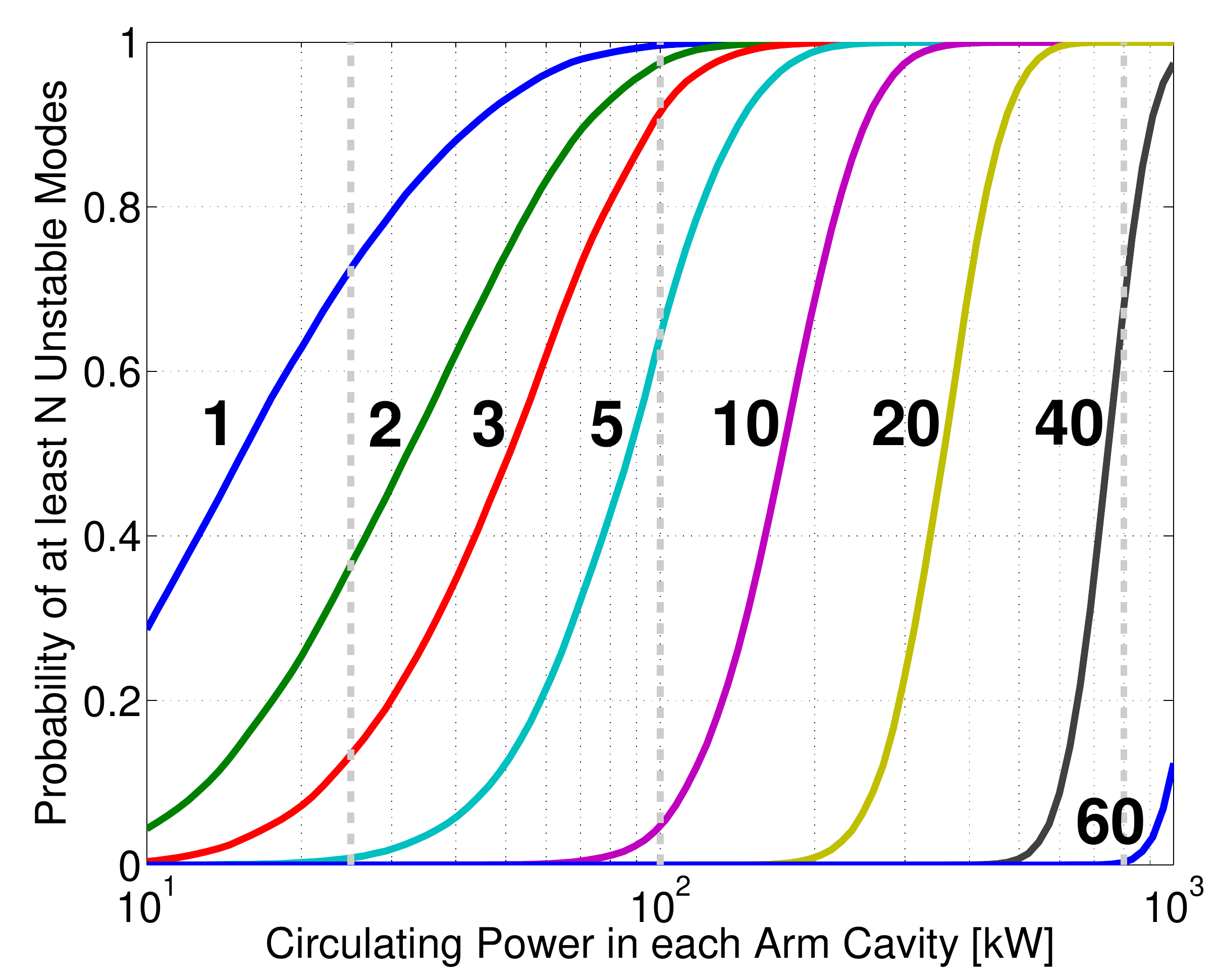}
\caption{The probabilities of having various numbers of unstable modes
 anywhere in the interferometer (with 4 test masses),
 as a function of circulating power, $\Parm$.
Each curve gives the probability that at least $N = \set{1, 2, 3, 5, \dots}$ unstable modes are observed in
 an Advanced LIGO interferometer at a given level of circulating power in the arm cavities.
The vertical grey lines mark the threshold power of the mode
 first observed to be unstable, near \SI{25}{kW} of circulating power,
 the power level at which the first observing run of Advanced LIGO is planned,
 roughly \SI{100}{kW}, and the design power level of \SI{800}{kW}.
These data indicate that more than 40 modes will likely need
 to be damped or otherwise defused in order to operate at \SI{800}{kW}.
\flabel{PI_ProbableNumber}}
\vspace{-8pt}
\end{figure}

\section{Conclusions}
\sslabel{Conclusions}

Parametric instabilities, studied in great depth and feared as a limitation to the
 attainable operating powers of interferometric \gw\ detectors,
 have been observed for the first time in an Advanced LIGO detector.
The behavior of the observed instability was found to be largely in agreement
 with models of the effect, implying that no significant ingredients have been omitted
 in the theoretical analysis.

Furthermore, the observed PI has been quenched by thermally tuning the optical resonance
 of the interferometer away from the resonance of the associated mechanical mode.
This approach to PI, while sufficient for a small number of potentially unstable modes,
 may not be sufficient at higher operating powers where many modes are available for
 runaway excitation.

Thanks to many years of theoretical and experimental work on parametric instabilities,
 now informed by the observations described in this paper,
 the challenge faced by high-power interferometric \gw\ detectors is clear and well understood.
While the necessary mitigation techniques are not trivial,
 a suitable combination of thermal tuning, active damping of excited mechanical modes,
 and passive reduction of mechanical mode Q-factors is expected to be sufficient to allow
 Advanced LIGO to operate stably at full power.

\vspace{10pt}
\begin{acknowledgments}
The authors would like to acknowledge the extensive theoretical analysis of
 parametric instabilities by our Moscow State University colleagues
 Vladimir Braginsky, Sergey Strigin, and Sergey Vyatchanin,
 without which these instabilities would have come as a terrible surprise.
 
LIGO was constructed by the California Institute of Technology and
 Massachusetts Institute of Technology with funding from the National Science Foundation,
 and operates under cooperative agreement PHY-0757058.
Advanced LIGO was built under award PHY-0823459.
This paper carries LIGO Document Number LIGO-P1400254.
\end{acknowledgments}



\fixme{some broken references!}

\bibliography{/Users/mevans/Documents/Work/Latex/papers}

\end{document}

%% file: mevansTex.tex
\usepackage[pdftex]{color}
\usepackage[pdftex]{graphicx}
\usepackage{amsfonts}
\usepackage{amsmath}
\usepackage{bm}                 

\newcommand{\sslabel}[1]{\label{sec:#1}}
\newcommand{\elabel}[1]{\label{eqn:#1}}
\newcommand{\flabel}[1]{\label{fig:#1}}

\newcommand{\cref}[1]{\ref{cpt:#1}}

\newcommand{\eref}[1]{\ref{eqn:#1}}
\newcommand{\fref}[1]{\ref{fig:#1}}

\newcommand{\listBegin}{\begin{tabular}{cp{4.5in}}}
\newcommand{\listEnd}{\end{tabular}}

\newcommand{\matrixBegin}[1]{\left[\!\!\left[ \begin{array}{#1}}
\newcommand{\matrixEnd}{\end{array} \right]\!\!\right]}

\newcommand{\beq}[1]{\begin{equation}\elabel{#1}}
\newcommand{\eeq}{\end{equation}}

\newcommand{\beqa}[1]{\begin{eqnarray}\elabel{#1}}
\newcommand{\eeqa}{\end{eqnarray}}

\newcommand{\set}[1]{\left\{ #1 \right\}}

\def\({\left(}
\def\){\right)}
\def\[{\left[}
\def\]{\right]}

\def\gw{gravitational wave}

\def\third{3\textsuperscript{rd}}

\usepackage{siunitx}

\DeclareSIUnit\rtHz{$\sqrt{\si{\Hz}}$}
\DeclareSIUnit\mrtHz{\meter\per\rtHz}
\DeclareSIUnit\msolar{{$M_\odot$}}

\definecolor{spring}{rgb}{0.7,0.9,0.7}
\definecolor{brick}{rgb}{0.7,0.2,0.1}
\definecolor{redHL}{rgb}{1.0,0.5,0.5}

\newcommand{\fixme}[1]{\colorbox{redHL}{#1}}

%% file: abbrev.tex
\def\gw{gravitational wave}

